\documentclass[12pt]{iopart}

\usepackage{cite}

\usepackage{graphicx}
\usepackage{dcolumn}

\begin{document}

\title[On the Hellmann potential]{On an approximation to the Schr\"odinger equation with the Hellmann potential}

\author{Paolo Amore\dag \ and Francisco M Fern\'andez
\footnote[2]{Corresponding author}}

\address{\dag\ Facultad de Ciencias, Universidad de Colima, Bernal D\'iaz del
Castillo 340, Colima, Colima, Mexico}\ead{paolo.amore@gmail.com}

\address{\ddag\ INIFTA (UNLP, CCT La Plata-CONICET), Divisi\'on Qu\'imica Te\'orica,
Blvd. 113 S/N,  Sucursal 4, Casilla de Correo 16, 1900 La Plata,
Argentina}\ead{fernande@quimica.unlp.edu.ar}

\maketitle

\begin{abstract}
We tried to determine the range of validity of a recently proposed
modification of the Hellmann potential that leads to analytical
eigenvalues and eigenfunctions.  We discuss the difficulties that
we found in the analysis of the main equations and results. We
conclude that the eigenvalues reported by the authors do not
exhibit the same order as those of the Hellmann potential thus
leading to a different underlying physics. What is more: the
spectrum of the modified model is qualitatively different from the
one supported by the Hellmann potential.
\end{abstract}

\section{Introduction}

Some time ago, Hellmann\cite{H35} proposed an approximation to the study of
atoms in which the atomic kernel is treated by means of the Thomas-Fermi
equation and the valence electrons by means of the Schr\"{o}dinger one. In
this way the author derived a simple potential for the valence electrons of
the form $V(r)=-1/r+(A/r)e^{-2\kappa r}$ in atomic units. This potential
also proved suitable for the study of metallic binding\cite{HK36}.

In a recent paper Arda and Server\cite{AS14} obtained approximate
expressions for the eigenvalues and eigenfunctions of the Hellmann potential
as well as for a non-Hermitian variant. They resorted to a suitable
modification of the Coulomb interaction and the centrifugal part of the
radial eigenvalue equation in order to obtain an exactly solvable equation.
Since the authors did not discuss the range of validity of the substitutions
carried out we tried to fill this gap.

\section{The approach}

The authors studied the Schr\"{o}dinger equation with the Hellmann potential
\begin{equation}
V(r)=\frac{-a+be^{-\lambda r}}{r}  \label{eq:V(r)}
\end{equation}
The behaviour near the origin is given by $V(r)\approx (b-a)r^{-1}$ and at a
great distance from the origin by $V(r)\approx -a/r$. Therefore, if $a>0$
the attractive Coulomb tail at sufficiently large $r$ supports an infinite
number of bound states.

The radial part of the Schr\"{o}dinger equation is
\begin{equation}
\left\{ \frac{d^{2}}{dr^{2}}-\frac{l(l+1)}{r^{2}}+\frac{2m}{\hbar ^{2}}%
\left[ E-V(r)\right] \right\} R(r)=0  \label{eq:radial}
\end{equation}
where $l=0,1,\ldots $ is the angular-momentum quantum number and the
boundary conditions
\begin{eqnarray}
&R(0)=0&  \nonumber \\
\lim\limits_{r\rightarrow \infty }&R(r) =0&  \label{eq:BC}
\end{eqnarray}
apply to the bound states.

In order to solve the equation analytically the authors carried out the
following substitutions
\begin{equation}
\frac{1}{r}\rightarrow \frac{\lambda }{1-e^{-\lambda r}},\;\frac{1}{r^{2}}%
\rightarrow \frac{\lambda ^{2}}{\left( 1-e^{-\lambda r}\right) ^{2}}
\label{eq:approximations}
\end{equation}
that have already been used by other authors in the past\cite{D64,M83,BB88}.
It follows from the Taylor expansions
\begin{eqnarray}
\frac{\lambda }{1-e^{-\lambda r}} &=&\frac{1}{r}+\frac{\lambda }{2}\,+\frac{{%
\lambda }^{2}r}{12}\,-\ldots   \nonumber \\
\frac{\lambda ^{2}}{\left( 1-e^{-\lambda r}\right) ^{2}} &=&\frac{1}{r^{2}}+%
\frac{\lambda }{r}+{\frac{5{\lambda }^{2}}{12}}\,+\frac{{\lambda }^{3}r}{12}%
\,+\ldots   \label{eq:expanded_approximations}
\end{eqnarray}
that the errors increase with $\lambda $. The purpose of this paper is to
estimate the effect of these substitutions on the spectrum of the model.

By means of the ansatz
\begin{equation}
\psi (u)=R(-\ln (u)/\lambda )=u^{\lambda _{1}}(1-u)^{\lambda _{2}}F(u)
\label{eq:psi(u)}
\end{equation}
Arda and Sever obtained a differential equation for $F(u)$. Since the
transformation $u=e^{-\lambda r}$ maps $0\leq r<\infty $ onto $1\geq u>0$
the exponents $\lambda _{1}$ and $\lambda _{2}$ should be positive in order
to have a solution that satisfies both boundary conditions. In order to
obtain a suitable differential equation the authors chose
\begin{eqnarray}
\lambda _{1}^{2} &=&\frac{2m}{\lambda ^{2}\hbar ^{2}}(E+a\lambda )+l(l+1)
\nonumber \\
\lambda _{2} &=&\frac{1}{2}\left[ 1\pm \sqrt{1+4l(l+1)}\right]
\label{eq:lambda_j}
\end{eqnarray}
but, curiously, did not specify the sign of these parameters. As argued
previously both should be positive.

The authors stated that the function $F(u)$ reduces to a polynomial when
\begin{eqnarray}
-n &=&\lambda _{1}+\lambda _{2}+\frac{1}{2}\sqrt{-\frac{8m}{\lambda
^{2}\hbar ^{2}}(E+b\lambda )}  \nonumber \\
n &=&0,1,\ldots  \label{eq:quantization}
\end{eqnarray}
Since the left- and right-hand sides of this equation have opposite signs
one concludes that there is no possible solution. However, from this
equation the authors derived the following expression for the energy
\begin{eqnarray}
E &=&-\frac{m}{8\hbar ^{2}(n+l+1)^{2}}\left\{ 4(a^{2}+b^{2})\right.
\nonumber \\
&&+4\frac{\hbar ^{2}}{m}\lambda b\left[ 2l^{2}+(n+l)^{2}+l(3+2n)\right]
\nonumber \\
&&+\frac{\lambda ^{2}\hbar ^{4}}{m^{2}}\left[ l(1+2n)+(n+l)^{2}\right] ^{2}
\nonumber \\
&&+\left. 4a\left[ -2b+\frac{\lambda \hbar ^{2}}{m}\left[
l(1+2n)+(n+l)^{2}\right] \right] \right\}  \label{eq:E}
\end{eqnarray}
that we rewrote in such a way that $m$ and $\hbar $ always appear in the
ratio $\hbar ^{2}/m$. In their table~1 the authors gave dimensionless values
to the potential parameters $a$, $b$, and $\lambda $ but did not specify the
ratio $\hbar ^{2}/m$. In order to test this expression they chose $\lambda
=0 $, $\hbar =1$ and obtained the hydrogenic energy levels
\begin{equation}
E=-\frac{ma^{2}}{2(n+l+1)^{2}}
\end{equation}
but never specified the value of $m$. The authors made their paper even more
unclear when in their table 1 chose $n\geq l+1$ instead of the quantum
number indicated above in equation (\ref{eq:E}). When $n=l=0$ the energy
given by equation (\ref{eq:E}) becomes independent of $\lambda $ in
disagreement with the results in their table~1.

We could not reproduce the authors' results in the third column of their
table~1 by trying some reasonable choices of $\hbar ^{2}/m$ in the their
expression (\ref{eq:E}). To make any analysis even more difficult the
authors claimed to compare their results in table 1 with those of the
references [11] and [28] of their paper (present references \cite{H35} and
\cite{HC08}). However, those references do not show any result for the
eigenvalues of the Hellmann potential.

Some time ago Adamowski\cite{A85} calculated the eigenvalues of the
Hamiltonian with the Hellmann potential
\begin{equation}
H=-\frac{\hbar ^{2}}{2m}\nabla ^{2}-\frac{A}{r}+\frac{Be^{-Cr}}{r}
\label{eq:H(A,B,C)}
\end{equation}
where $A,C>0$. This paper, which is useful for present purposes as shown
below, was omitted by Arda and Sever\cite{AS14}. It follows from the
Hellmann-Feynman theorem
\begin{equation}
\frac{\partial E}{\partial C}=-B\left\langle e^{-Cr}\right\rangle
\label{eq:HF}
\end{equation}
that
\begin{eqnarray}
E(C &\rightarrow &\infty )<E<E(C=0),\;B>0  \nonumber \\
E(C &=&0)<E<E(C\rightarrow \infty ),\;B<0  \label{eq:Bounds}
\end{eqnarray}
where $E$ is the energy of any bound-state solution to $H\psi =E\psi $.
These bounds can be calculated exactly because the potential is Coulombic at
both limits\cite{A85}.

Adamowski chose the length and energy units $a_{0}=\hbar ^{2}/(mA)$ and $%
mA^{2}/(2\hbar ^{2})$, respectively. After separation of the angular part of
the Schr\"{o}dinger equation the remaining radial equation becomes
\begin{equation}
\left\{ -\frac{d^{2}}{dr^{2}}+\frac{l(l+1)}{r^{2}}-\frac{2}{r}+\frac{b}{r}%
e^{-\lambda r}\right\} R(r)=ER(r)  \label{eq:radial_2}
\end{equation}
where $b=2B/A$ and $\lambda =a_{0}C$. In this case the bounds derived above
are
\begin{eqnarray}
E(\lambda &=&0)=-\frac{(2-b)^{2}}{4\nu ^{2}},\;\nu =1,2,\ldots  \nonumber \\
E(\lambda &\rightarrow &\infty )=-\frac{1}{2\nu ^{2}}  \label{eq:bounds_2}
\end{eqnarray}

Now that we have an equation derived in a clear way we are able to calculate
its eigenvalues. Table~\ref{tab:E} shows the eigenvalues of the radial
equation (\ref{eq:radial_2}) with $\lambda =0.01$ and $b=1$. From left to
right the four columns display the states labelled as in the hydrogen atom,
present results obtained by means of the Riccati-Pad\'{e} method\cite{FMT89a}%
, the results of Adamowski\cite{A85}, and those of Arda and Sever\cite{AS14}%
. Although we do not know the value of $m/\hbar ^{2}$ chosen by the latter
authors or how they calculated their eigenvalues the column of results
labelled \textit{present} in their table~1 seems to match the other two ones
quite satisfactorily. At first sight the approximate eigenvalues reported by
Arda and Sever appear to agree reasonably well with those calculated
accurately for equation (\ref{eq:radial_2}).\ However, the order of the
almost degenerate energy levels appears to be incorrect. If we denote the
energy levels by $E_{\nu \,l}$, where $\nu =n+l+1=1,2,\ldots $ is the
principal quantum number, then we appreciate that the eigenvalues of (\ref
{eq:radial_2}) exhibit the order $E_{\nu \,l+1}<E_{\nu \,l}$ whereas the
substitution (\ref{eq:approximations}) leads to the opposite order (assuming
that the results of Arda and Sever were already calculated for the Hellmann
potential with the substitutions indicated by equation (\ref
{eq:approximations})).

The Riccati-Pad\'{e} method enables us to calculate the eigenvalues with
much more accuracy than the one in table~\ref{tab:E}. We do not show more
accurate results here because it is obviously unnecessary for present
purposes.

Our original purpose of determining the range of validity of the
substitutions (\ref{eq:approximations}) was hindered by the fact that we
could not reproduce the results of the authors' table 1 with the authors'
analytical expression (\ref{eq:E}). However, we can easily show that the
approximation is bound to fail for sufficiently large values of $\lambda $.
The reason is that the modified Coulomb potential $-\frac{2\lambda }{%
1-e^{-\lambda r}}$ does not longer support an infinite number of bound
states and, what is more, this number shrinks to none at some critical value
of $\lambda $. The addition of the Hulthen-like potential $\frac{b\lambda
e^{-\lambda r}}{1-e^{-\lambda r}}$ does not change this fact.

\section{Conclusions}

It has been quite difficult to determine the range of validity of the
approximation proposed by Arda and Sever\cite{AS14} for several reasons. In
the first place, it is not clear to us how they derived their expression for
the bound states. In the second place, this expression does not appear to
yield the eigenvalues shown in their table~1. In the third place, they
failed to indicate the references reporting the eigenvalues used for
comparison. We adopted the point of view that such inconsistencies are
merely due to misprints and typos and compared their expression for the
energy eigenvalues obtained by other approaches. Our analysis suggests that
the eigenvalues obtained by the authors for a given $\nu $ and different $l$
exhibit the wrong order; in other words: the analytical formula describes
the underlying physics incorrectly. In addition to it, the discrepancy
between the eigenvalues calculated with the left and right expressions in (%
\ref{eq:approximations}) should increase with $\lambda $. This is probably
the reason why the authors only showed results for quite small values $%
\lambda $. Our analysis above shows that the substitution (\ref
{eq:approximations}) transforms a problem with an infinite number of bound
states into one with a finite number. As $\lambda $ increases the bound
states of the modified model disappear one by one until a critical value is
obtained beyond which there is no eigenvalue. This drastic change introduced
by the substitution (\ref{eq:approximations}) was entirely omitted by the
authors.

\ack F. M. Fern\'andez would like to thank the University of Colima for
financial support and hospitality.

\begin{table}[]
\caption{Eigenvalues of the radial equation (\ref{eq:radial_2}) with $%
\lambda=0.01$ and $b=1$.}
\label{tab:E}
\begin{center}
\par
\begin{tabular}{lD{.}{.}{11}D{.}{.}{6}D{.}{.}{6}}
\hline \multicolumn{1}{c}{State}& \multicolumn{1}{c}{Present} &
\multicolumn{1}{c}{Ref.\cite{A85}} & \multicolumn{1}{c}{Ref.\cite{AS14}} \\
\hline
1s &  -0.2598520035  &  -0.25985   &    -0.26502  \\
2s &  -0.07192801595 &  -0.07193   &    -0.07760  \\
2p &  -0.07202032438 &  -0.07202   &    -0.07502  \\
3s &  -0.03656400027 &  -0.03656   &    -0.04300  \\
3p &  -0.03664789365 &  -0.03664   &    -0.04180  \\
3d &  -0.03681429863 &  -0.03681   &    -0.03947  \\
4s &  -0.02363657974 &  -0.02364   &    -0.03102  \\
4p &  -0.02371070818 &  -0.02371   &    -0.03031  \\
4d &  -0.02385702542 &  -0.02386   &    -0.02891  \\
4f &  -0.02407191089 &  -0.02407   &    -0.02690  \\

\end{tabular}
\par
\end{center}
\end{table}

\end{document}